# Hazard recognition in an immersive virtual environment: Framework for the simultaneous analysis of visual search and EEG patterns

**Mojtaba Noghabaei, SM.ASCE[1] and Kevin Han, Ph.D., M.ASCE[2]**

[1]Department of Civil, Construction, and Environmental Engineering, North Carolina State University, P.O. Box 7908, Raleigh, NC 27695-7908; e-mail: snoghab@ncsu.edu

[2]Department of Civil, Construction, and Environmental Engineering, North Carolina State University, P.O. Box 7908, Raleigh, NC 27695-7908; e-mail: kevin_han@ncsu.edu

**ABSTRACT**

Unmanaged hazards in dangerous construction environments proved to be one of the main sources of injuries and accidents. Hazard recognition is crucial to achieve effective safety management and reduce injuries and fatalities in hazardous job sites. Still, there has been lack of effort that can efficiently assist workers in improving their hazard recognition skills. This study presents virtual safety training in an Immersive Virtual Environment (IVE) to enhance worker's hazard recognition skills. A worker wearing a Virtual Reality (VR) device, that is equipped with an eye-tracker, virtually recognizes hazards on simulated construction sites while a brainwave-sensing device records brain activities. This platform can analyze the overall performance of the workers in a visual hazard recognition task and identify hazards that need additional intervention for each worker. This study provides novel insights on how a worker's brain and eye act simultaneously during a visual hazard recognition process. The presented method can take current safety training programs into another level by providing personalized feedback to the workers.

## INTRODUCTION

The construction industry has been reported more than 60,000 fatal injuries throughout the world annually (Jeelani et al. 2017b; Lingard 2013). Particularly, fatal construction injuries in the US alone have increased by more than 16% from 2011 to 2015 (BLS 2015). The associated costs of such events surpass $48 billion in the US(Jeelani et al. 2017b) and reduce the profit margin of the construction projects. Safety-related accidents negatively affect construction projects' success and endanger the financial stability of minor companies (Zou and Sunindijo n.d.). Studies show that construction workers' hazard recognition inability and ineffective hazard management cause poor construction safety performance (Albert et al. 2014). For instance, more than 42% of construction incidents are related to worker-related issues, such as workers' poor hazard recognition skills(Haslam et al. 2005).

Hazard recognition significance for construction safety is fully recognized in construction industry practice and research (Jeelani et al. 2019a). Hazard recognition is generally considered as the key component of any injury prevention program (Carter and Smith 2006). If workers efficiently manage and recognize hazards, injuries can be significantly reduced (Jeelani et al. 2019a). Studies indicate that over 57% of job site hazards remain unrecognized by the construction workers (Bahn 2013; Carter and Smith 2006; Jeelani et al. 2019a; Perlman et al. 2014). Workers' hazard recognition skills can be improved by adopting new safety training programs by the construction industry (Li et al. 2015; Rozenfeld et al. 2010).

Recent studies suggest utilizing new technologies such as Virtual Reality (VR) (Zhao and Lucas 2015), eye-tracking (Jeelani et al. 2019c), and brain-sensing in safety training programs to detect cognitive and physiological behaviors of workers during visual hazard recognition tasks. Moreover, recent studies illustrate that employing eye-tracking and VR technologies in safety training programs can outstandingly enhance workers' hazard recognition skills (Jeelani et al. 2017c). Despite these improvements, the workers are still failing to efficiently recognize hazards while contributing to safety training programs (Perlman et al. 2014). Although the safety training programs can be helpful to some levels, additional enhancements can be done. For instance, Jeelani et al. (2017a) suggested using eye-tracking for improving safety training programs; however, they did not consider using VR to reduce workers' exposure to hazards. Also, they did not consider using brain sensors to investigate workers' brain patterns during visual hazard recognition. This limitation can be addressed by utilizing brain sensors and VR. (Jeelani et al. 2017c)also suggested virtually simulated construction sites using VR to improve safety training outcomes since VR provides higher spatial perception than conventional approaches, such as 2D videos or books. Also, this study failed to utilize eye trackers and brain sensors during a visual hazard recognition process. In addition to VR and eye-tracker, brain wave sensors empower researchers to collect brain wave signals during hazard recognition tasks and classify brain activity patterns. This classification can help researchers to accurately understand how workers' brain react to a certain type of hazard. Also, researchers recognized that workers' emotion deviates significantly while working in a hazardous environment and the researchers were able to perceive workers' stress levels by obtaining electroencephalography (EEG) signals. Merging these studies and technologies altogether can potentially improve current safety training programs and open up a new field of study for safety researchers.

## STUDY OBJECTIVES

To improve current safety research and address previous research limitations, the authors have proposed a framework to integrate VR, brain-sensing, and eye-tracking technologies using a VR head-mounted device (HMD) with eye-tracking capability and an EEG sensor. This framework helps to understand workers' brain and eye patterns during a visual hazard recognition process while providing immersive experience using VR. The recorded data from the eye-tracker and brain sensor can later be analyzed and classified using a machine learning technique to recognize the patterns in eye movement and brain activities. To achieve this platform, several issues must be solved. One of these challenges is synchronization of the data in the same platform as the eye-tracking and brain sensors have high-frequency sampling rates. In addition, the authors have to justify based on previous research that the data generated by this platform is reliable since the integration of EEG and VR HMD can potentially introduce artifacts for both EEG and eye-tracking data. Finally, the authors have designed a hazard recognition experiment to record data, synchronize EEG and eye-tracking data, and provide customized feedback to the participants.

The presented platform can take current safety training to the next level by providing customized feedback to workers and identifying hazard types that need additional interventions. This paper describes the proposed platform and the following research steps:

(1) introduction to the platform sections,

(2) an experiment including data collection from participants who detected hazards on a virtually simulated construction site,

(3) synchronize EEG and eye-tracker data, and

(4) detect the type of hazards that need additional intervention according to the detection rate by the experiment.

## BACKGROUND

Hazard recognition is an important contributor to effective construction safety management (Dzeng et al. 2016). Researchers have stated that existing safety management programs were ineffective since managers and workers were unable to recognize hazards in job sites (Tam et al. 2004). To avoid injuries, it is vital to recognize safety hazards and adopt injury prevention techniques at construction sites (Carter and Smith 2006; Noghabaei et al. 2020). In cases where safety hazards remain unidentified, workers are inclined towards dangerous acts, which eventually leads to workers' unexpected exposure to harm, and tragic injuries (Albert et al. 2014). To improve current safety training, researchers have developed a safety training using VR (Jeelani et al. 2017c). The results indicate that safety training programs that use VR provide more precise simulations for the workers. Hence, these programs can drastically improve the outcomes of safety training. Overall, VR can provide better spatial perception than traditional visualization techniques such as 2D screens. Therefore it can help in improving the quality of training (Balali et al. 2018; Kayhani et al. 2019; Noghabaei et al. 2019). Also, researchers have proposed a framework that employs 360-degree panoramas recorded videos from real construction sites and used the videos in VR training. the outcomes indicate that the proposed framework will significantly improve workers' hazard recognition skills (Eiris et al. 2018).

Pedram et al. (2017) evaluated the VR safety training frameworks and illustrated that these frameworks have a significant positive learning experience. Furthermore, construction researchers proposed combining VR with EEG to measure humans' responses in virtually designed spaces (Ergan et al. 2019). On the other hand, eye-tracking and a combination of eye-tracking and brain-sensing can present exceptional insights about how the workers recognize hazards in a VR environment. Researchers have demonstrated that localizing gaze fixation can predict the likelihood of detecting a hazard in a work zone (Jeelani et al. 2019b). A combination of EEG, eye-tracking, and VR technologies in a single platform can take safety training to a new level; however, integration of all the devices in a single platform introduces several challenges such as synchronization of EEG and eye-tracking. This paper solves the challenges associated with developing such a framework and provides proof of concept for a framework that combines, brain sensors, eye-trackers, and VR.

## METHOD

To achieve the objectives of this study, an experiment was defined accordingly. In this experiment, a participant recognizes hazards in a VR simulated construction site. Meanwhile, a brain sensor and eye tracker record eye movements and brainwaves of the participant. This paper only discusses the synchronization and detected hazard types. A detailed analysis of the signals will be discussed in future research.

**Simulated Environment**

The first step of this research is to simulate a construction site in a VR environment. To make VR simulations, the authors utilized a 3D engine, Unity 3D, that is commonly used in the architectural, engineering, and construction (AEC) industry. Studies suggested that human behavior is similar in an immersive virtual environment in comparison with physical built environments (Heydarian et al. 2015). Therefore, VR provides a unique environment for assessing human behaviors (Heydarian et al. 2015). Ten hazards in Table 1. are simulated in the virtual construction site. Figure 1(A) demonstrates a first-person-view of the simulated site. Figure 1(A) and (C) show fall hazards that are Hazard 1 and Hazard 4. Ten introduced hazards in Table 1. List of hazards in the virtual environment accounts for 80% of construction hazards that caused fatalities (Helander 1991; Jeelani et al. 2018).

**Table 1. List of hazards in the virtual environment**

| Hazard id | Description |
|---|---|
| 1 | Fall hazard: unprotected object near the edge |
| 2 | Electrical hazard: unprotected electric cables without proper conduit |
| 3 | Trip hazard: unprotected ladder |
| 4 | Fall hazard: unprotected barrel near the edge |
| 5 | Chemical hazard: an unmarked bucket with unknown chemical fluid without lid |
| 6 | Trip hazard: unprotected bricks on the ground |
| 7 | Electrical hazard: unprotected junction box without proper protection |
| 8 | Chemical hazard: unprotected igneous chemical fluids |
| 9 | Chemical hazard: an unmarked bucket with unknown chemical fluid without lid |
| 10 | Pressure hazard: gas cylinder without proper restraints in the work zone |

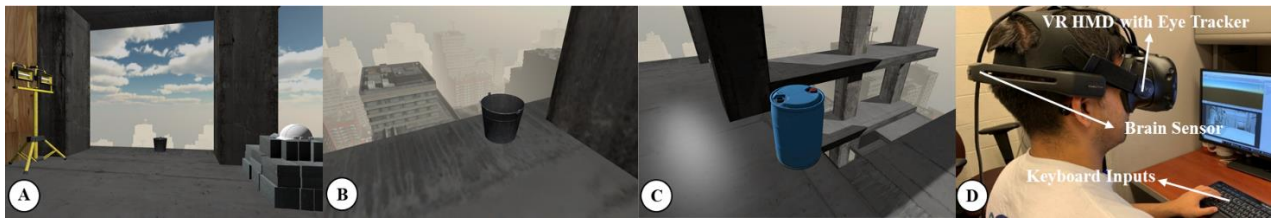

**Figure 1. 3D simulations; (a) simulated construction site; (b) hazard number one; (c) hazard number four; (d) a participant wearing EEG and VR HMD**

**Data Acquisition**

EMOTIV EPOC+ is used to record EEG data. It is a commonly used brain sensor device by researchers. This device has 14 channels and has been widely used in construction research for investigating workers' brain waves. To record eye-tracking data in VR, the HTC Vive Pro Eye VR headset is utilized. This HMD was released in 2018 as the best eye-tracking enabled VR headset produced by joint combination the best research-grade eye-tracking manufacturer (Tobii) and best VR HMD manufacturer (HTC). Figure 1(D) shows a participant wearing both EEG and VR HMD.

**Data Synchronization**

One of the goals of this study is the synchronize EEG recordings with the eye-tracking data. Synchronizing EEG and eye-tracking data is challenging due to the high frequency and different sampling rates of the devices and latency issues of computer programs. Based on the literature, there are three approaches to synchronize EEG and eye-tracking data (Dimigen et al. 2011). The first approach presents the use of a shared trigger. In this method, common trigger pulses are sent from the computer to both eye-tracking and EEG devices. A y-shaped wire connects the computer to both EEG and eye-tracking devices. The main benefit of this method is that the same signal is used for synchronization; therefore, the synchronization accuracy is quite high. However, this approach is not always possible due to hardware limitations (e.g. most EEG and eye-tracker devices don't allow hardware manipulation). The second approach is to insert short text strings in eye-tracking data when triggers are sent to the EEG device. These messages are used later to synchronize data. This method is hardware-independent and can provide relatively high accuracy. The last approach is to use an analog output. In this approach, eye-tracking data is fed directly into the EEG device. A digital to analog converter card of the eye-tracker outputs the data as an analog signal. This signal can be directly fed into the EEG device. Although this approach affords easy synchronization, it requires hardware manipulation, which cannot be done in many EEG and eye-tracking devices.

In this study, messages and event markers were used to synchronize the data (Figure 2). Recordings from eye-tracking and EEG data were synchronized later using the EYE-EEG toolbox (Dimigen et al. 2011; "EYE-EEG" 2018). This toolbox uses common trigger pulses and messages sent from the stimulation PC (running both Unity 3D and Emotive software) to both systems. Once a button is pressed by participants, a developed wrapper code that works based on windows raw input API sends inputs to both Unity and Emotive software. This wrapper hooks native input events and allows receiving inputs even when the Emotive software or Unity applications are working in the background. This code helps to receive the raw input events at the same time by both Unity and Emotive to reduce any latency issues. Afterward, Unity sends messages to the eye-tracking device, and Emotive software sends events markers to the EEG device. These messages and Event markers are recorded in the data stream. These recordings are used later by EYE-EEG to synchronize EEG data and eye-tracking data. This approach is a very accurate approach and it is commonly used for synchronization of EEG and eye-tracking data (Meyberg et al. 2015, 2017)

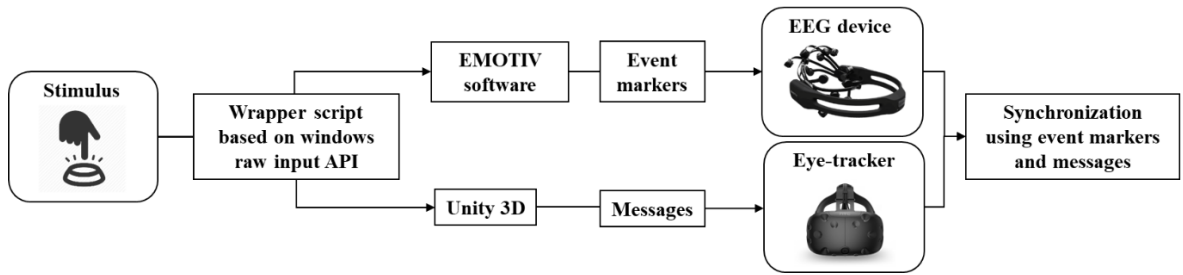

**Figure 2. synchronization overview**

# EXPERIMENTAL SETUP
## Device Calibration

The data was gathered from 10 participants who had different levels of understanding of hazard recognition. The average age of the participants was 25.1. All participants were familiar with VR and previously used VR HMDs. To ensure that the participants are familiar with construction hazard recognition, a very brief introduction to safety in construction were given to the participants. The introduction contained information about what is considered a hazard. There was no history of mental disorder or any eye-related problem. Each participant had 10 trials with a one-minute rest between each trial. To reduce errors related to the sequence effect (learning effect) (Wang et al. 2017), hazard locations are changed in each trial. In this experiment, the learning effect means the affected brain signals due to previous trials in the experiment. In addition, each trial was limited to 30 seconds to make sure that the participants are focusing on hazard recognition task during the experiment. Prior to performing the experiment, the eye-tracking device is calibrated by asking the participants to look at the red dots in the VR simulation (5-point calibration) as instructed by the manufacturer of the eye-tracker. Then, EEG calibration was performed as instructed by the manufacturers of the EEG device. To select proper labels (i.e., identified hazards and non hazards) for the data, the participants were asked to press a controller button as they detect the hazards. Furthermore, EYE-EEG is utilized to synchronize EEG data and eye-tracking data (Meyberg et al. 2015, 2017) as previously mentioned.

# EXPERIMENTAL RESULTS
## Safety Related Results

Table 2 demonstrates the detection rate of all participants for each hazard. A hazard was labeled as identified where a participant pressed the button while he was looking at the hazard within a one-second time period before pressing the button. For example, 20.65% of all recognized hazards by all participants were related to hazard number 1. Table 2 provides important insights into the future direction of personalized safety training.

| Table 2. The detection rate of all participants for different hazards | | | | | | | | | | |
|---|---|---|---|---|---|---|---|---|---|---|
| Hazard id | 1 | 2 | 3 | 4 | 5 | 6 | 7 | 8 | 9 | 10 |
| Detection Ratio | 20.65% | 8.18% | 5.77% | 6.99% | 5.58% | 6.05% | 9.97% | 6.33% | 11.63% | 18.86% |

## DISCUSSIONS
**Industry Implications**

Figure 3 combines the detection ratio for five major hazard categories. Figure presents important insights about which hazard types require intervention and more attention.

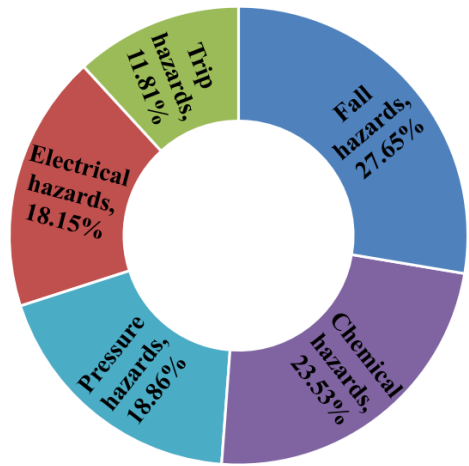

**Figure 3. The ratio of major hazard types that were detected by participants.**

Based on the results of this study, the participants incline to focus more on fall and chemical hazards. On the other hand, the participants were not able to identify trip hazards. For instance, the detection ratio for unprotected ladder (Ha 3) in Table 2 was only about 5%, however this type of hazard (unprotected ladder) accounts for around 10% of construction injuries and fatalities (Helander 1991). In addition, participants detection ratio for the gas cylinder (Ha 10) in Table 2 was around 19%, but in real construction sites, only 5% of the total construction fatalities were related to this type of hazard. These results suggest the importance of focusing on fatal hazards rather than focusing on all hazards at the same time. In the future, researchers should focus on teaching the workers how to prioritize hazards types. In this way, the construction industry can potentially save money while reducing the number of fatalities and injuries in workplaces.

**Study Limitations and Future Works**

Despite the contributions and strengths of this research, it has shortcomings to be addressed in the future. Although this environment was presented in an immersive environment with high-quality standards, it can be improved significantly using high definition physically based rendering techniques. Using this technique, VR simulations can be indistinguishable from reality which leads to more accurate EEG and eye-tracking recordings. In addition, future research can focus on comparing the brain waves and eye movements of the participants in the virtual environments vs. physical-built environments and define metrics on how accurate an environment was simulated.

Furthermore, to achieve high-quality signals, artifact removal is necessary (Sherafat et al. 2019; Taghaddos et al. 2016). Developing and validating noise cancellation is necessary for this framework. More importantly, this research did not investigate hazard types and how they affect human brains differently. Further investigations can be done to understand what types of hazards stimulate which parts of the brain. In addition, it is possible to detect which types of hazards causing more arousal in the brain. Finally, the combination of this platform with a camera and robotic systems can improve safety by active safety prevention (Asadi et al. 2019a; b; Asadi and Han 2018).

## CONCLUSIONS

Blending visual search and brain wave analyses provides valuable information for safety trainers and educators. Based on the results of this study, a brain sensor, eye-tracker, and VR can be combined and synchronized in a single platform. The findings of this paper provide three main directions for future research. First, analyzing EEG and eye-tracking signals and classify the signals based on whether a worker detected a hazard. Secondly, identifying EEG channel locations that correspond to hazard recognition. Lastly, using this framework for improving human-computer interaction in the construction industry.

## ACKNOWLEDGEMENT

This work is partially funded by the National Institute for Occupational Safety and Health (NIOSH, T42-OH008673). Any opinions, findings, conclusions, or recommendations presented in this paper are those of the authors and do not reflect the views of NIOSH the individuals acknowledged above.